\documentclass[preprint,12pt]{elsarticle}

\usepackage{graphicx}

\usepackage{amssymb}

\usepackage{tikz}
\usepackage{ulem}

\usepackage{wasysym}
\usepackage{eucal}
\usepackage{amsfonts}
\usepackage{lineno}

\journal{Journal Name}

\begin{document}

\begin{frontmatter}


\title{Monitoring L\'evy-Process Crossovers}



\author{Maike A. F. dos Santos$^1$, Fernando D. Nobre$^1$, 
and Evaldo M. F. Curado$^1$}

\address{$^1$Centro Brasileiro de Pesquisas F\'isicas and National Institute of Science and Technology for Complex Systems, Rua Xavier Sigaud 150, Rio de Janeiro, RJ 22290-180, Brazil}

\begin{abstract}
The crossover among two or more types of diffusive processes represents a vibrant theme in nonequilibrium 
statistical physics. 
In this work we propose two models to generate crossovers among different L\'evy processes:
in the first model we change gradually the order of the derivative in the Laplacian term of the diffusion equation, 
whereas in the second one we consider a combination of fractional-derivative diffusive terms 
characterized by coefficients that change in time. 
 The proposals are illustrated by considering semi-analytical 
(i.e., analytical together with numerical) procedures to follow the time-dependent 
solutions. We find changes between two different regimes and it is shown that, far from the crossover regime, both models yield 
qualitatively similar results, although these changes may occur in 
different forms for the two models.  
The models introduced herein are expected to be useful for describing 
crossovers among distinct diffusive regimes that occur frequently in complex systems.  
\end{abstract}


\end{frontmatter}


\newpage
\section{Introduction}
\label{S:1}

According to Einstein's theory, the problem of Brownian motion may be 
seen as a diffusion of Brownian particles through a homogeneous environment.
Consequently, it follows the linear diffusion equation~\cite{reichlbook,rednerbook}, 

\begin{eqnarray}
\frac{\partial p(x,t)}{\partial t} = D \, \frac{\partial^{2} p(x,t)}{\partial x^{2}}~,   
\label{eqlin-diffusion}
\end{eqnarray}

\noindent
where $D$ is the diffusion constant and $p(x,t)$ represents a continuous 
probability density for finding the tagged particle in a position between $x$ 
and $x+dx$, at time $t$
(for simplicity, we restrict ourselves to a one-dimensional space).  
The well-known solution of this equation is the Gaussian distribution, 

\begin{equation}
p(x,t) = {1 \over (4 \pi Dt)^{1/2}} \exp [-x^{2}/(4Dt)]~, 
\label{eqgauss-dist}
\end{equation}

\noindent
so that the mean-square displacement, 

\begin{equation}
\langle (\Delta x)^2 \rangle = 2Dt~, 
\label{eqx2linear}
\end{equation}

\noindent
increases linearly with $t$, a typical signature of the Brownian motion and all 
linear-diffusion processes. 

However, nowadays a wide variety of systems has presented non-Gaussian 
distributions~\cite{reichlbook,rednerbook,tsallis1997levy,tsallisbook}, which may 
be found as natural consequences of several theoretical studies, like 
nonextensive statistical mechanics~\cite{tsallisbook,tsallisreview}, 
superstatistics~\cite{sposini2018random,chechkin2017brownian,slkezak2018superstatistical}, diffusion with memory kernels~\cite{dos2018non,dos2019fractional,hristov2019response}, controlled diffusion~\cite{dos_Santos_2019,PhysRevE.68.046121,dosSantos2019}, rare events~\cite{rebenshtok2016complementary,wang2018renewal}, and nonlinear Fokker-Planck 
equations~\cite{schwammle2008q,plastino95,tsallis96,schwammle2009dynamics,%
curado2018equilibrium,casas2019nonlinear}. 
To take into account non-Gaussian distributions, essentially two distinct procedures 
have been mostly proposed in the literature, for   
generalizing the linear-diffusion equation in Eq.~(\ref{eqlin-diffusion}),  
as described next. 
(i) In the first approach, one keeps the standard derivatives and 
introduces nonlinearities in the 
probability $p(x,t)$; the most known case leads to the porous-medium equation, 
where one introduces a power in the probability of the diffusion 
term~\cite{tsallisbook,schwammle2008q}. 
By adding an extra contribution in the porous medium equation, due 
to an external confining potential,  
one obtains a nonlinear Fokker-Planck
equation~\cite{plastino95,tsallis96,schwammle2009dynamics,%
curado2018equilibrium,casas2019nonlinear}. These types of nonlinear
equations present solutions known in the literature as $q$-Gaussian distributions
[which recover the Gaussian of Eq.~(\ref{eqgauss-dist}) 
in the limit $q \to 1$], being intimately related to nonextensive statistical 
mechanics, leading to a wide variety of applications in social, economical, 
physical, and biological 
systems~\cite{tsallisbook,tsallisreview,schwammle2008q,%
plastino95,tsallis96,schwammle2009dynamics,%
curado2018equilibrium,casas2019nonlinear,troncoso2007}.
(ii) In the second procedure, one keeps the probability linear, but turn the 
standard derivatives into fractional
derivatives~\cite{metzler2000random,metzler2014anomalous,dos2019analytic}. 
More particularly, by introducing a fractional derivative in the diffusion term,
the corresponding solutions are known as L\'evy 
distributions~\cite{levy1954theorie,zaburdaev2015levy,bertoin1996levy,%
nolan2003stable,gnedenko1968limit,tsallisprl1995}.  
In the present investigation we restrict ourselves to this second approach; 
according to this later scenario,
the non-Gaussian distributions are connected with anomalous-diffusion processes, 
through a power-law behavior in the mean-square displacement, i.e. $
\langle (\Delta x)^2 \rangle = 2 \mathcal{K}_{\alpha} t^{\alpha}$ $(\alpha>0)$, where 
$\mathcal{K}_{\alpha}$ represents a general diffusion coefficient with fractional dimension. 
Actually, such relation is usually associated with various diffusive behaviors~\cite{dos2019analytic,metzler2014anomalous}, 
classified as  sub- ($0<\alpha<1$), normal- ($\alpha=1$), super- ($1<\alpha<2$), and hyper- 
($\alpha>2$) diffusive processes; the particular case $\alpha=2$ is known as ballistic diffusion.

Many studies in the literature were dedicated to L\'evy statistics~\cite{levy1954theorie,metzler2000random}, and the so-called 
L\'evy distributions, which became powerful tools to approach some complex systems~\cite{zaburdaev2015levy,bertoin1996levy}.
The most general form of a L\'evy distribution is written in terms of two indexes, $L_{\mu,\beta}(z)$, where $\mu$ is a real exponent
(to be defined below) and $\beta$ represents an asymmetry parameter~\cite{rednerbook}; herein, we will deal with symmetric L\'evy 
distributions, corresponding to $\beta=0$. Hence, for simplicity, we define $L_{\mu}(z) \equiv L_{\mu,0}(z)$, expressed in terms of 
the following Fourier transform,     

\begin{eqnarray}
L_{\mu}(z) = \displaystyle
 \frac{1}{2\pi}\int_{-\infty}^{\infty}  e^{i k z  -a|k|^{\mu}} dk~. 
\label{levyevolution}
\end{eqnarray}

\noindent
The integral above is nonnegative for $0<\mu \leq 2$ and consequently it may be 
considered as a probability
density only in this range~\cite{reichlbook};  it recovers the Gaussian and 
Cauchy distributions in the particular limits
$\mu=2$ and $\mu=1$, respectively.
For $0<\mu < 2$, these distributions present power-law 
tails, $L_{\mu}(z) \sim z^{-(1+\mu)}$, leading to 
a divergence in the mean-square displacement, 
$\langle (\Delta z)^2 \rangle \to \infty$~\cite{tsallis1997levy,nolan2003stable}. 

From the theoretical point of view, the L\'evy distribution appears in a generalized version of central-limit theorem, 
where the sum of identically distributed random variables follows power laws, instead of 
a Gaussian distribution~\cite{gnedenko1968limit,tsallisprl1995}. 
In the scenario of the fractional Fokker-Planck equation~\cite{metzler2000random}, the fractional diffusion  
can be constructed through the Scher-Montroll equation~\cite{montroll1965random,scher1975anomalous,montroll1973random},
where a particular case is expressed in terms of the Riesz derivative, presenting 
the L\'evy distribution as a solution. 
The Riesz fractional derivative has found applications that go beyond statistical mechanics, 
e.g., in 
a fractional Schr\"odinger equation~\cite{laskin2002fractional} that appears as a potential model 
to explain some laser measurements~\cite{longhi2015fractional,barthelemy2008levy}.
Moreover, the L\'evy distributions were found in fractional-collision models (Rayleigh, driven Maxwell 
gas)~\cite{PhysRevE.68.055104,barkai2004stable}, 
models with single big jumps~\cite{PhysRevE.100.012108}, cold atoms trapped in optical 
lattices~\cite{aghion2017large}, and stochastic resetting processes~\cite{PhysRevE.92.052127}.

Nowadays the fractional dynamics (in Riesz sense) has become an useful tool for describing the protein sliding 
in polymeric systems (e.g., DNA)~\cite{lomholt2005optimal}. In these situations the occurrence of a single big jump 
is associated with the protein folding. Thereby, the dynamics for a single type of 
L\'evy process can be defined by a continuous 
probability $p(x,t)$ following

\begin{eqnarray}
\frac{\partial \, p(x,t)}{\partial t} = \frac{\partial^{\mu} \, p(x,t)}{\partial |x|^{\mu}}~.  
\label{eq2levy}
\end{eqnarray}

\noindent
One should notice that the equation above
suggests a scaling between position and time, i.e., $|x| \sim t^{1/\mu}$; therefore, the 
solution of Eq.~(\ref{eq2levy}) is expressed in terms of a L\'evy distribution,  

\begin{eqnarray}
p(x,t)=\frac{1}{t^{\frac{1}{\mu}}}
L_{\mu}\left[ \frac{|x|}{t^{\frac{1}{\mu}}} \right]~. 
\label{levyevolution}
\end{eqnarray}

It is well known that protein walks in DNA play an important role in reparation of some 
genetic sequences~\cite{blainey2006base} 
that include mutations or chromosomal defects.
In this scenario, the dynamic behavior of proteins in DNA can be described by a search-for-target model~\cite{lomholt2005optimal}; 
this model includes a fractional Laplacian term (like in Eq.~(\ref{eq2levy})) to take into account the big jumps among different 
parts of DNA folds. 
Within this context, one can think of the possibility for a given walker to be governed 
by two different L\'evy processes in distinct time regimes. 
Similar situations, exhibiting crossovers between anomalous to normal diffusion processes, have been observed    
in telomeres in the nucleus of mammalian cells~\cite{PhysRevLett.103.018102}, 
diffusion in biological cells~\cite{hofling2013anomalous}, as well as in various complex  
systems~\cite{molina2018crossover,PhysRevResearch.1.023006,PhysRevE.76.021111,spagnolo2008volatility,mantegna1994stochastic,dos2019mittag}. 

However, a theoretical support for crossovers between different L\'evy processes 
remains open. 
In the present work we investigate this question by proposing two models that may 
exhibit such kind of phenomenon, presenting the ability to generate crossovers 
among different L\'evy processes 
described through the fractional Riesz derivative. 
In section~\ref{sec2} we introduce the two models: 
(i) In the first one we change gradually the order of the derivative in the Laplacian term of the diffusion equation;
(ii) In the second one we consider a combination of fractional-derivative diffusive terms 
characterized by coefficients that change in time. 
 In both cases we work out analytical time-dependent solutions; then, we
illustrate typical crossovers by following numerically the time evolution of the probability distribution. 
It is shown that, far from the crossover regime, both models yield qualitatively 
similar results, although the 
changes may occur in different forms for the two models. 
Finally, in section~\ref{conclusion} we present ou main conclusions and discuss 
possible applications for the models introduced. 

\section{Models for crossovers between L\'evy processes} 
\label{sec2}

Let us consider a random walk, exhibiting a crossover in time,
described by a normalized continuous probability, 
  
\begin{eqnarray}
\int_{-\infty}^{\infty} dx \, p(x,t) =1 \qquad (\forall t),  
\label{normalizationcond}
\end{eqnarray}

\noindent
following L\'evy processes according to Eq.~(\ref{eq2levy}). 
The crossover is signalled by two different exponents 
$\mu_{1}$ ($t \ll t_{\rm cross})$ and $\mu_{2}$ ($t \gg t_{\rm cross})$, 
where $t_{\rm cross}$ represents a 
typical time for the change between the two regimes to occur. 
Next we propose two models that produce crossovers among L\'evy flights. 
The first one considers the 
fractional derivative in the  Laplacian term of the diffusion equation with a variable 
order, whereas the second one
is defined by two fractional-derivative diffusive terms, each of them multiplied 
by different time-dependent diffusion coefficients.  

\subsection{Model 1: Time-Dependent Order of the Fractional Derivative}

The idea of fractional-differential operators with variable order was introduced 
recently~\cite{lorenzo2002variable,lorenzo2007initialization}, 
leading to an increasingly interest and various applications, like 
anomalous diffusion~\cite{sun2009variable,yang2017new}, memory modelling~\cite{sun2011comparative}, 
reaction-diffusion problems~\cite{hajipour2019accurate}, wave equation~\cite{heydari2019computational}, 
telegraph equation~\cite{gomez2019time,hosseininia2019meshfree}, and other 
investigations~\cite{almeida2019variable,ortigueira2019variable,DABIRI201840,FU201537,sun2019review,Gorenflo2005}. 

Herein, we adopt the following time-dependent-order Riesz 
fractional-derivative definition for a two-variable function $f(x,t)$~\cite{zhuang2009numerical,lin2009stability}, 

\begin{eqnarray}
\frac{\partial^{\mu(t)} f(x,t)}{\partial |x|^{\mu(t)}} = b[\mu(t)] \int_0^{+\infty}d\xi \ \frac{f(x+\xi,t)-2 f(x,t)-f(x-\xi,t)}{\xi^{1+\mu(t)}}~,
\label{lin}
\end{eqnarray}

\noindent
where $b[\mu(t)] = \frac{\Gamma(1+\mu(t))}{\pi}\sin\left(\frac{\mu(t) \pi}{2}\right)$ and $1 \leq \mu(t) \leq 2$. 
A numerical analysis of the definition above was carried in Ref.~\cite{lin2009stability}, showing its stability and 
convergence  in the diffusion equation. 
The Fourier transform of this derivative is expressed as 

\begin{eqnarray}
\mathcal{F} \left \lbrace  \frac{\partial^{\mu(t)}  \ \ \ }{\partial |x|^{\mu(t)}} f(x,t) \right\rbrace = -|k|^{\mu(t)}f(k,t)~,
\label{13}
\end{eqnarray}

\noindent
with $f(k,t)=\int_{-\infty}^{\infty} dx f(x,t)e^{- i k x}$.

In this scenario we modify Eq.~(\ref{eq2levy}) by introducing a time-dependent-order 
fractional derivative
according to Eq.~(\ref{lin}),  

\begin{eqnarray}
\frac{\partial}{\partial t}p(x,t)=\frac{\partial^{\mu(t)} \ \ \ }{\partial |x|^{\mu(t)}} p(x,t)~.  
\label{variabeLevy}
\end{eqnarray}

\noindent
The initial and boundary conditions are the usual ones, 

\begin{eqnarray}
p(x,0)=\delta(x)~;  \qquad \qquad p(x\rightarrow \pm \infty, t)=0~, 
\label{conditions}
\end{eqnarray}

\noindent
and taking into account the purpose of the present work, i.e., a crossover between 
two different regimes, we
impose the following asymptotic limits for Eq.~(\ref{variabeLevy}), 

\begin{eqnarray}
\frac{\partial}{\partial t}p(x,t)&=& \frac{\partial^{\mu_1} \ \ \ }{\partial |x|^{\mu_1}} p(x,t) \quad  (t\rightarrow 0), 
\label{asymptzero} \\
\frac{\partial}{\partial t}p(x,t)&=& \frac{\partial^{\mu_2} \ \ \ }{\partial |x|^{\mu_2}} p(x,t) \quad  (t\rightarrow \infty).
\label{asymptinfty}
\end{eqnarray}

Now, using Eq.~(\ref{13}), we may perform the Fourier transform of 
Eq.~(\ref{variabeLevy}) to obtain

\begin{eqnarray}
\frac{\partial}{\partial t}p(k,t)&=& -|k|^{\mu(t)} p(k,t),
\end{eqnarray}

\noindent
implying on

\begin{eqnarray}
p(x,t)=\frac{1}{2\pi}\int_{-\infty}^{+\infty}dk \exp\left [-\int_0^{t}|k|^{\mu(t')}dt' \right ] e^{i k x}~, 
\label{solutionmodel1}
\end{eqnarray}

\noindent
which recovers the  L\'evy-distribution solution for  $\mu$ constant. 

 Given a form for $\mu(t)$, the integral above may be computed 
for the time interval $[0,t]$ by employing standard numerical tools. Herein, we 
followed the time evolution of the distribution in Eq.~(\ref{solutionmodel1}) by making 
use of the pack~\textit{NIntegration} from a symbolic computational software, 
which plots the numerical solution $p(x,t)$ for chosen times.  

As can be noticed, there are multiple
forms for $\mu(t)$ that interpolate between the limits given by 
Eqs.~(\ref{asymptzero}) and~(\ref{asymptinfty}). Herein, we have taken into account two 
ingredients, common in natural systems, for defining these forms: 
(i) They should present a time-scale parameter (to be defined below as $\tau$), 
so that the different regimes may be identified by $t \ll \tau$ and $t \gg \tau$;
(ii) They should contemplate some typical situations for the crossover to occur, 
namely, abruptly and smoothly.
Hence, to illustrate crossovers 
between different regimes, we propose below 
two distinct recipes for $\mu(t)$,
interpolating between the Gaussian ($\mu=2$) and L\'evy ($1 \leq \mu < 2$) regimes. 
The first one is given by 

\begin{eqnarray}
\mu(t)= 2\theta \left( 1-  {t \over \tau} \right) + \frac{\Gamma+1}{2} 
\left(1+ e^{-\frac{t}{\tau}} \right) \theta \left( {t \over \tau} - 1 \right)~,
\label{firstcase}
\end{eqnarray}

\noindent
where $\Gamma$ is a real parameter in the range from unit to 
$(3-e^{-1})/(1+e^{-1}) \approx 1.924$ (i.e.,  $\Gamma \in [1,1.924]$).  
 Due to the Heaviside function $\theta(y)$, this proposal is expected to 
cover abrupt changes between the two regimes; the second one (which
should cover smooth changes) is defined by 

\begin{eqnarray}
\mu(t)= 2 - \gamma e^{-\frac{t}{\tau}}~,  
\label{secondcase}
\end{eqnarray}

\noindent
where $\gamma$ is a real parameter ($\gamma \in (0,1]$). In both cases  
$\tau$ plays the role of a relaxation time, and herein it will be related,
in most cases, to the crossover time, i.e., $t_{\rm cross} \propto \tau$. 
Although these proposals are only illustrative, the
changes between the two limiting behaviors, 
$\mu_{1}$ ($t \ll \tau)$ and $\mu_{2}$ ($t \gg \tau)$, 
occur more abruptly (smoothly) in Eq.~(\ref{firstcase}) [Eq.~(\ref{secondcase})],
thus covering two distinct physical situations. 

\begin{figure}[h!]
\centering
\includegraphics[scale=0.70]{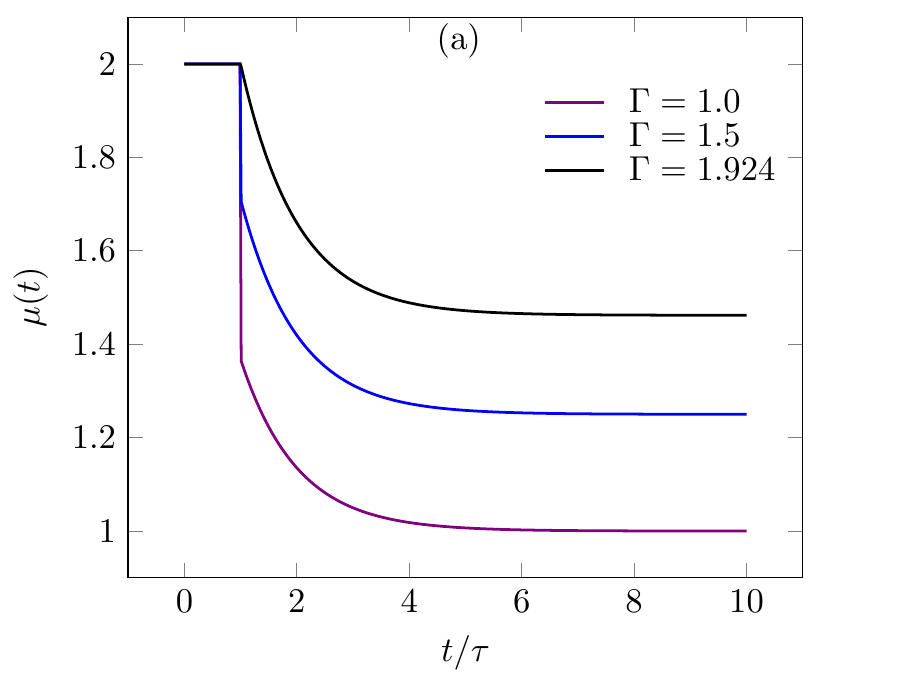}  
\includegraphics[scale=0.70]{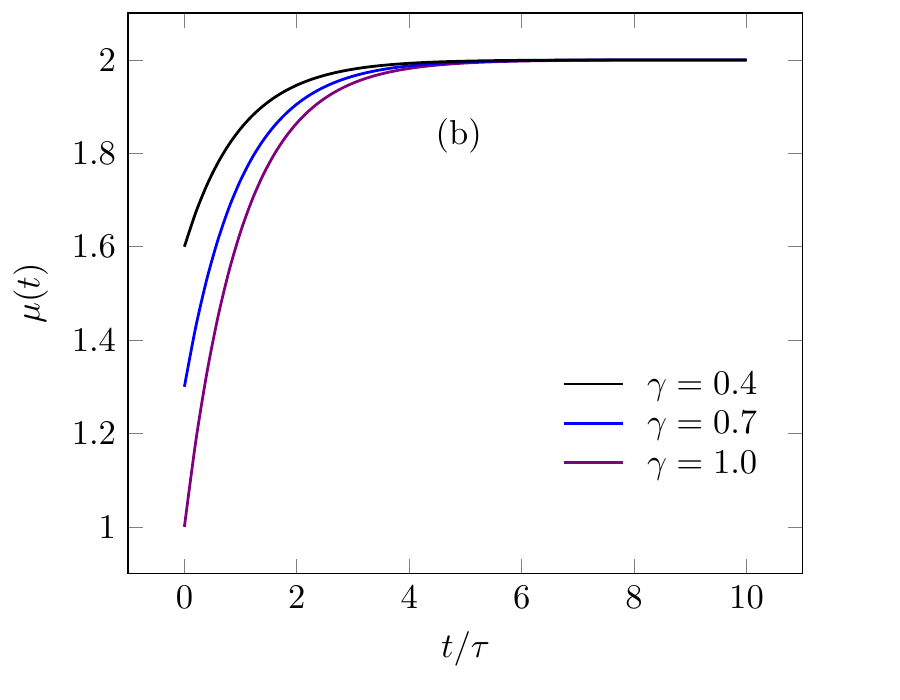} 
\caption{\small{ 
The time-dependent order of the fractional 
derivative $\mu(t)$ [cf. Eq.~(\ref{variabeLevy})]
is represented versus time (conveniently rescaled by the crossover time $\tau$):
(a) Following Eq.~(\ref{firstcase}) for typical values of the parameter $\Gamma$
(increasing values of $\Gamma$ from bottom to top); 
(b) Following Eq.~(\ref{secondcase}) for typical values of $\gamma$
(increasing values of $\gamma$ from top to bottom). 
}}
\label{fig1}
\end{figure}

The time evolution of $\mu(t)$ is shown in Fig.~\ref{fig1} for both cases of  
Eq.~(\ref{firstcase}) [Fig.~\ref{fig1}(a)] and 
Eq.~(\ref{secondcase}) [Fig.~\ref{fig1}(b)]. 
In Fig.~\ref{fig1}(a) one finds crossovers between the Gaussian
[$\mu(t)=2$,  for $t < \tau$] and typical L\'evy [$\mu(t)=(\Gamma+1)/2$,  
for $t \gg \tau$] regimes. 
On the other hand, in Fig.~\ref{fig1}(b) crossovers between distinct L\'evy 
regimes [$\mu(t)=2-\gamma$, for $t < \tau$] to the Gaussian
[$\mu(t)=2$,  for $t \gg \tau$] are exhibited. 
Although in both cases the crossovers in $\mu(t)$ take place typically in the time 
interval $\tau < t < 5\tau$, so that for $t > 5\tau$ the final regime has been 
fully attained, the changes in Fig.~\ref{fig1}(a) are more 
abrupt than those found in Fig.~\ref{fig1}(b), as expected. However, 
as will be shown next, crossovers in the corresponding distributions
require much larger times to occur. 

\begin{figure}[h!]
\centering
\includegraphics[scale=0.70]{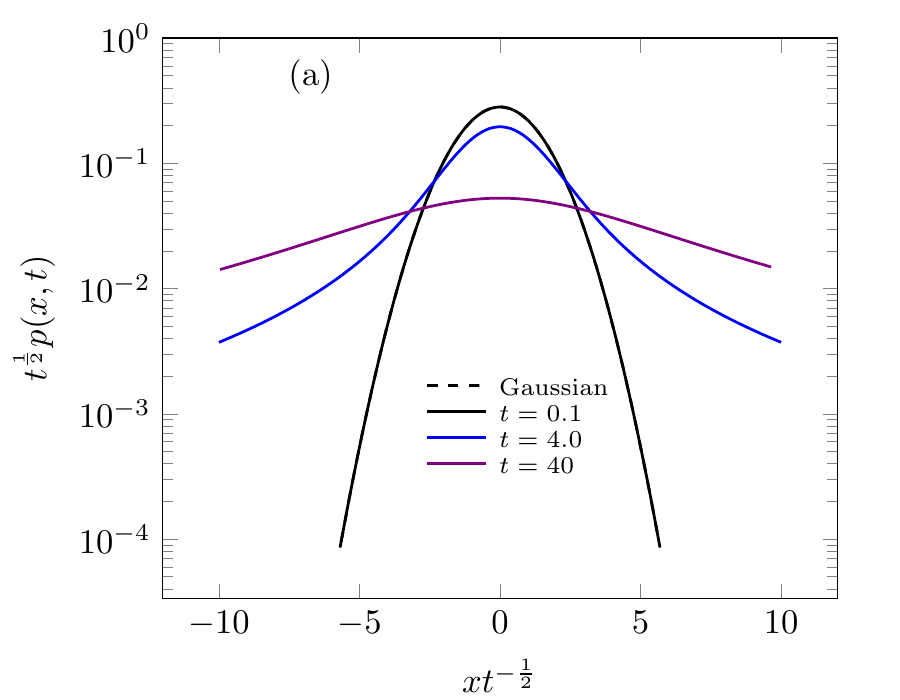} 
\hspace{0.2cm} 
\includegraphics[scale=0.70]{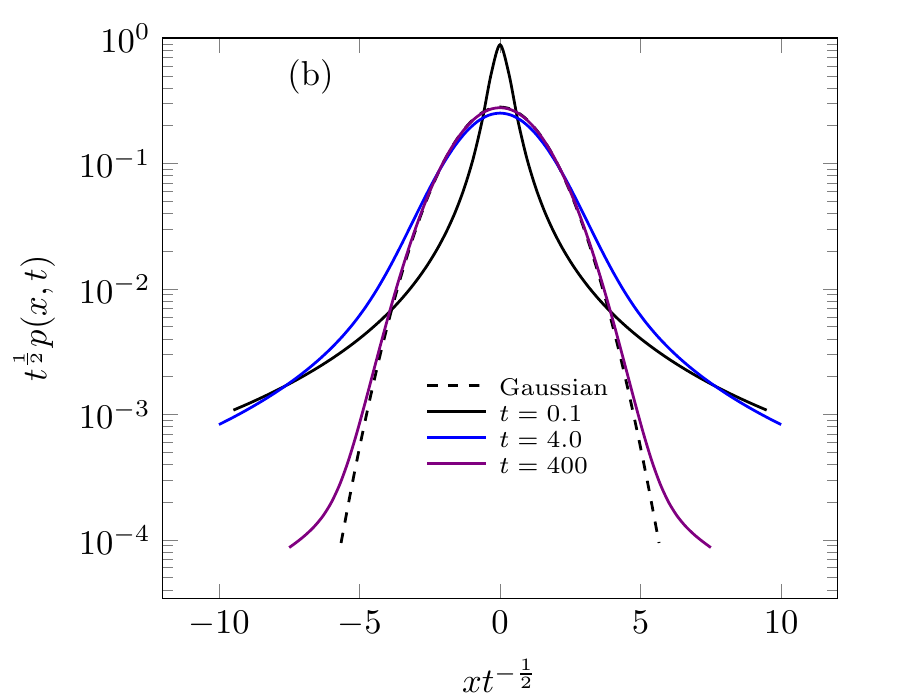} 
\caption{\small{
The probability distributions $p(x,t)$ are exhibited (using conveniently rescaled variables) 
in log-linear representations, for typical times. 
(a) $\mu(t)$ following Eq.~(\ref{firstcase}) for $\Gamma=1$ and $\tau=10$, showing 
longer tails for increasing times; the case $t=0.1$ appears superposed with the Gaussian.  
(b) $\mu(t)$ following Eq.~(\ref{secondcase}) for $\gamma=1$ and $\tau=10$, exhibiting 
shorter tails for increasing times; the case $t=0.1$ is typically a Cauchy distribution.
}}
\label{fig2}
\end{figure}

 Analyzing Eq.~(\ref{eqgauss-dist}) one figures out a convenient 
way for plotting a Gaussian distribution, i.e., through a representation in the 
rescaled variables, $t^{1/2}p(x,t)$ versus $xt^{-1/2}$, which takes into account 
the following aspects: 
(i) Its norm is preserved; 
(ii) It becomes width-independent, so that Gaussian distributions 
characterized by different widths all collapse into a single curve.
Based on this, we use this representation in Fig.~\ref{fig2}, where 
a Gaussian distribution appears either as the initial [Fig.~\ref{fig2}(a)],
or long-time distribution [Fig.~\ref{fig2}(b)]. In these figures, curves for typical 
times are shown, where each curve is 
represented by considering its own time variable $t$. 
Therefore, in Fig.~\ref{fig2} 
we show $p(x,t)$ for increasing times, with the time-dependent 
order of the fractional derivative
following Eq.~(\ref{firstcase}) with $\Gamma=1$ [panel (a)], and 
Eq.~(\ref{secondcase}) with $\gamma=1$ [panel (b)], considering $\tau=10$ 
in both cases.
In Fig.~\ref{fig2}(a) one expects a crossover between the Gaussian ($t \ll \tau$) 
and Cauchy ($t \gg \tau$) distributions; although the times considered
($0 \leq t \leq 4\tau$) show clearly the crossover region (still far from the
long-time regime), one notices the spreading of the distribution 
towards the expected long-tailed Cauchy distribution. 
On the other hand, in Fig.~\ref{fig2}(b) we present the 
crossover between the Cauchy ($t \ll \tau$)
and Gaussian ($t \gg \tau$) distributions; one notices that the approach to 
the Gaussian distribution (represented by the dashed
curve) appears first around the central region of the distribution. 
Although the convergence to this limit distribution is expected, the disappearance 
of the long tails requires 
larger computational times, in such a way that the probability distribution 
at $t=40 \tau$ is still different from a Gaussian. 

\begin{figure}[h!]
\centering
\includegraphics[scale=0.70]{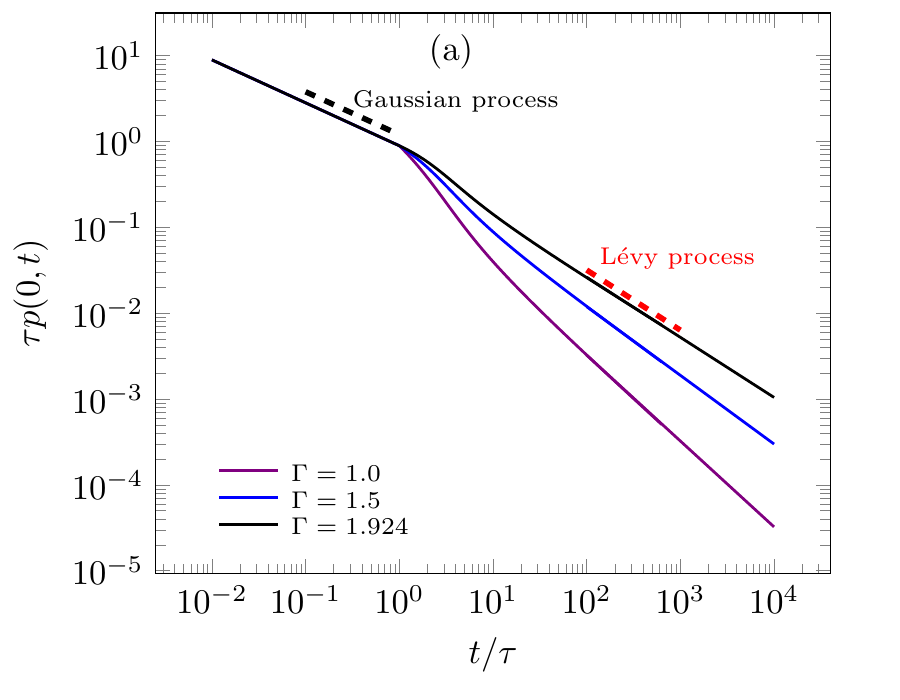} 
\hspace{0.2cm} 
\includegraphics[scale=0.70]{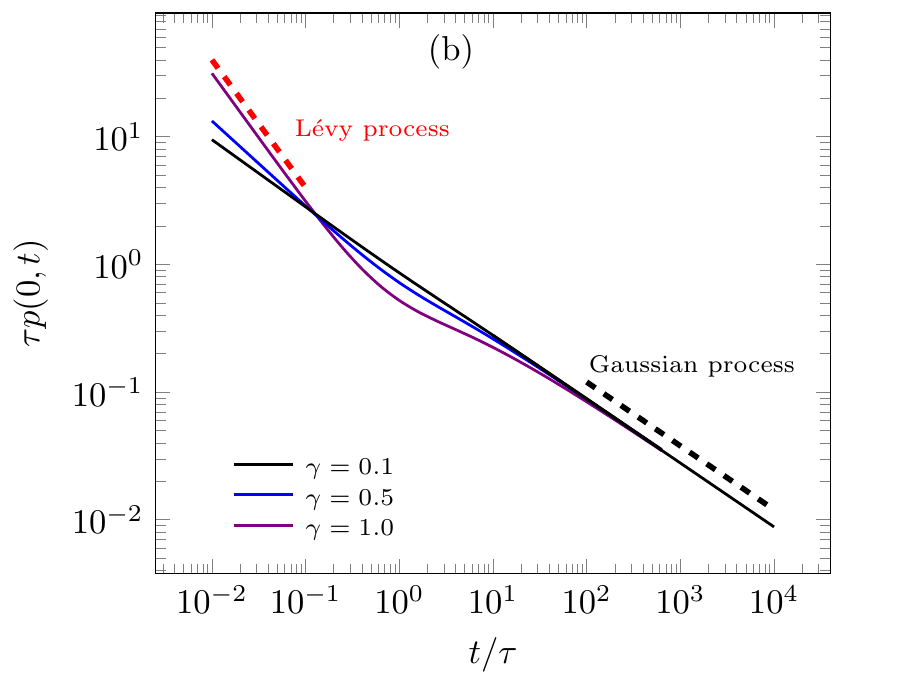} 
\caption{\small{
The return probabilities $p(0,t)$ are plotted versus time (using conveniently rescaled variables) 
in log-log representations, considering $\tau=10$.
(a) $\mu(t)$ following Eq.~(\ref{firstcase}) for typical values of the parameter
$\Gamma$ (in the L\'evy regime, increasing values of $\Gamma$ from bottom to top).
(b) $\mu(t)$ following Eq.~(\ref{secondcase}) for typical values of the parameter
$\gamma$ (in the L\'evy regime, increasing values of $\gamma$ from bottom to top).
Away from the crossover region, the scaling of Eq.~(\ref{returnprob}) 
is fulfilled, so that crossovers between the Gaussian ($\mu(t)=2$) to different 
L\'evy processes with $\mu(t)=(\Gamma + 1)/2$
are verified in (a), whereas crossovers between L\'evy processes with 
$\mu(t)=2-\gamma$ to $\mu(t)=2$ occur in (b). 
}}
\label{fig3}
\end{figure}

\begin{figure}[h!]
\centering
\includegraphics[scale=0.70]{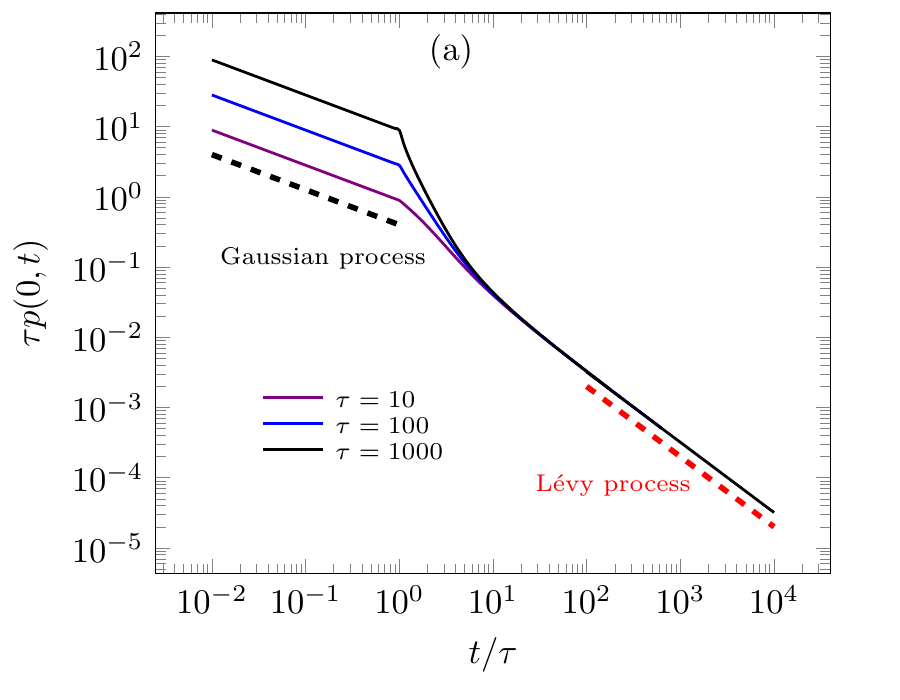} 
\hspace{0.2cm} 
\includegraphics[scale=0.70]{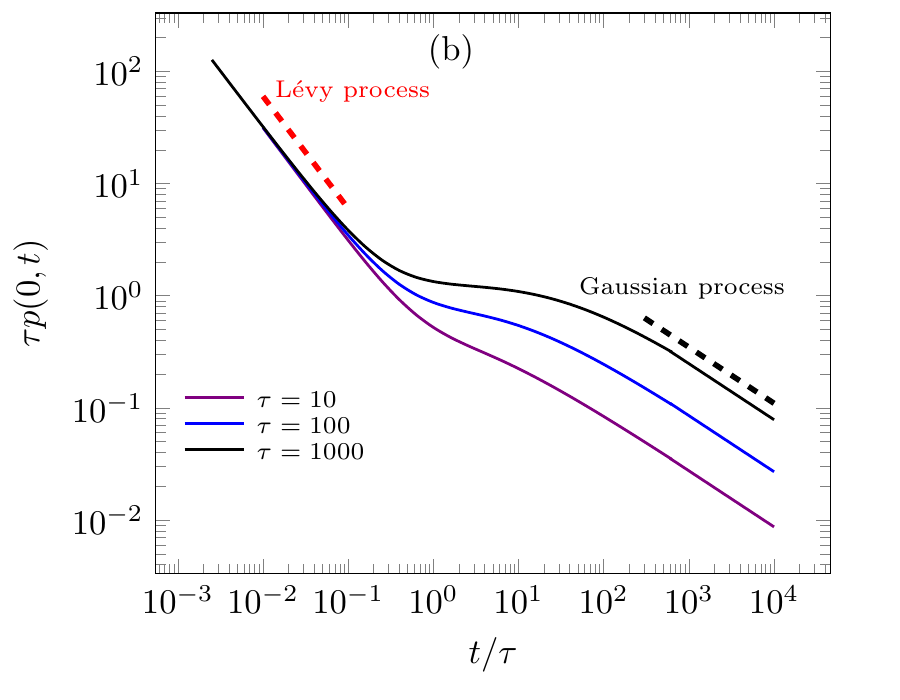} 
\caption{\small{
The return probabilities $p(0,t)$ are plotted versus time (using conveniently rescaled variables) 
in log-log representations for typical values of $\tau$.
(a) $\mu(t)$ following Eq.~(\ref{firstcase}) for $\Gamma=1$ (in the Gaussian regime, 
increasing values of $\tau$ from bottom to top).
(b) $\mu(t)$ following Eq.~(\ref{secondcase}) for $\gamma=1$ (in the Gaussian regime, increasing values of $\tau$ from bottom to top).
Away from the crossover region, the scaling of Eq.~(\ref{returnprob}) 
is fulfilled, so that crossovers between Gaussian ($\mu(t)=2$) to 
L\'evy ($\mu(t)=1$) processes
are verified in (a), whereas crossovers between L\'evy ($\mu(t)=1$) to 
Gaussian ($\mu(t)=2$) processes with occur in (b). 
}}
\label{fig4}
\end{figure}

The probability at the origin $p(0,t)$ (also known as return probability) has shown to provide important information
about L\'evy processes. In fact, according to Eq.~(\ref{levyevolution}) one has that 

\begin{eqnarray}
p(0,t) \sim t^{-1/\mu}~, 
\label{returnprob}
\end{eqnarray}

\noindent
in such a way that crossovers between distinct L\'evy regimes have been studied in the literature 
by making use of the above scaling for the return 
probability~\cite{mantegna1994stochastic,PhysRevE.52.1197,sokolov2004fractional,chechkin2008introduction}.  

Following this procedure, in Fig.~\ref{fig3} we exhibit the return probability $p(0,t)$ 
versus time in log-log representations, using conveniently rescaled variables, and
setting $\tau=10$.   
In Fig.~\ref{fig3}(a) we considered $\mu(t)$ following Eq.~(\ref{firstcase}), for  
typical values of $\Gamma$, whereas in Fig.~\ref{fig3}(b) we used 
$\mu(t)$ according to Eq.~(\ref{secondcase}) with different values of $\gamma$. 
These figures suggest that $\tau$ is related to a typical time required 
for a crossover to occur; moreover, away from the crossover region, 
the scaling of Eq.~(\ref{returnprob}) is fulfilled. One finds changes between 
the Gaussian ($\mu(t)=2$) to different L\'evy processes with $\mu(t)=(\Gamma + 1)/2$
in Fig.~\ref{fig3}(a), as well as between 
L\'evy processes with $\mu(t)=2-\gamma$ to $\mu(t)=2$ in Fig.~\ref{fig3}(b). 
In a similar way, in Fig.~\ref{fig4} we present $p(0,t)$ for several values of $\tau$, 
fixing $\Gamma=1$ [Fig.~\ref{fig4}(a)] and $\gamma=1$ [Fig.~\ref{fig4}(b)];  
these results reinforce the proposal that $\tau$ should be related 
to the crossover time, i.e., $t_{\rm cross} \propto \tau$. 
One observes crossovers between the Gaussian ($\mu=2$) 
to a Cauchy distribution ($\mu=1$)
in Fig.~\ref{fig4}(a), whereas those between the Cauchy and Gaussian 
distributions are found in Fig.~\ref{fig4}(b).
As expected, in both Figs.~\ref{fig3} and~\ref{fig4} one notices that the crossovers occur 
more abruptly in cases following Eq.~(\ref{firstcase}) [typically in the time intervals 
$1 < (t/\tau) < 10$ in Figs.~\ref{fig3}(a) and~\ref{fig4}(a)], than those of 
Eq.~(\ref{secondcase}) [typically in the time intervals
$10^{-1} < (t/\tau) < 10^{2}$ in Figs.~\ref{fig3}(b) and~\ref{fig4}(b)], thus covering 
two possible distinct real situations. 

 As will be discussed later, the model defined above should 
be applicable to processes presenting slow variations on its relevant environmental 
parameters, leading to smooth variations on the exponent $\mu(t)$. 
In such cases, intermediate L\'evy processes should appear, 
characterized by $\mu_{1} \leq \mu(t) \leq \mu_{2}$.  
As candidates, one mentions a new optical material constructed from glass
microspheres, so-called L\'evy glass~\cite{barthelemy2008levy}, light scattering in 
atomic vapours~\cite{Mercadier2013}, and cold atoms in 
optical lattices~\cite{Kessler2012}.

\subsection{Model 2: Time-Dependent Diffusion Coefficients}

The second model that we investigate consists in modifying Eq.~(\ref{eq2levy}) by 
introducing two competing diffusive 
contributions, characterized by different time-dependent diffusion coefficients, as follows

\begin{eqnarray}
\frac{\partial}{\partial t}p(x,t)=\mathcal{K}_1(t)\frac{\partial^{\mu_1 }}{\partial |x|^{\mu_1}} p(x,t)
+\mathcal{K}_2(t)\frac{\partial^{\mu_2 }}{\partial |x|^{\mu_2}} p(x,t)~.
\label{segundomodelo}
\end{eqnarray}

\noindent

To obtain crossovers between two L\'evy processes one may consider the 
above time-dependent diffusion 
coefficients leading to a prevalence of one of the fractional 
Laplacian terms at different  time regimes, e.g.,    
$\lim_{t \rightarrow 0} \mathcal{K}_1(t)= 1$ and  
$\lim_{t\rightarrow 0} [\mathcal{K}_2(t)/ \mathcal{K}_1(t)]= 0$, together with
$\lim_{t\rightarrow \infty} [\mathcal{K}_1(t)/ \mathcal{K}_2(t)]=0$ and 
$\lim_{t\rightarrow \infty} \mathcal{K}_2(t) = 1$. These choices lead to the same limits 
specified in Eqs.~(\ref{asymptzero}) and~(\ref{asymptinfty}), 
so that $p(x,t) \propto L_{\mu_1}(x,t)$ $(t\rightarrow 0)$, 
whereas $p(x,t) \propto L_{\mu_2}(x,t)$  $(t \rightarrow \infty)$.

Following the usual procedure, Eq.~(\ref{segundomodelo}) may be handled by performing a Fourier transform, 
leading to

\begin{eqnarray}
\frac{\partial \ }{\partial t }p(k,t)= -\left[ \mathcal{K}_1(t)|k|^{\mu_1} + \mathcal{K}_2(t)|k|^{\mu_2}\right] p(k,t)~,
\end{eqnarray}

\noindent
which admits the solution,

\begin{eqnarray}
p(k,t) = A(k) \varphi(k,t),
\end{eqnarray}

\noindent
where

\begin{eqnarray}
\varphi(k,t) =\exp\left [-|k|^{\mu_1}\int_0^t dt'\mathcal{K}_1(t') -|k|^{\mu_2}\int_0^t dt'\mathcal{K}_2(t') \right]~, 
\end{eqnarray}

\noindent
and $A(k)$ is an arbitrary function in Fourier space.  Using the conditions of Eq.~(\ref{conditions}) one obtains 

\begin{eqnarray}
p(x,t)=\frac{1}{2\pi}\int_{-\infty}^{+\infty}dk \, \varphi(k,t) e^{-i k x}~.
\end{eqnarray}

\noindent
In this way, the solution becomes a convolution of L\'evy distributions,

\begin{eqnarray}
p(x,t)=\frac{1}{\mathcal{P}_{\mu_1,\mu_2}(t)}\int_{-\infty}^{\infty}dx' 
L_{\mu_1}\left[ \frac{|x-x'|}{\left(\int_0^t dt'\mathcal{K}_1(t')\right)^{\frac{1}{\mu_1}}} \right]L_{\mu_2}\left[ \frac{|x'|}{\left(\int_0^t dt'\mathcal{K}_2(t')\right)^{\frac{1}{\mu_2}}} \right],
\label{solutionmodel2}
\end{eqnarray}

\noindent
where

\begin{eqnarray}
\mathcal{P}_{\mu_1,\mu_2}(t)=\left(\int_0^t dt' \mathcal{K}_1(t') \right)^{\frac{1}{\mu_1}} \left(\int_0^t dt'  \mathcal{K}_2(t') \right)^{\frac{1}{\mu_2}}.
\end{eqnarray}

 As done previously in the solution of Eq.~(\ref{solutionmodel1}), 
we used the pack~\textit{NIntegration} to follow 
the time evolution of the distribution in Eq.~(\ref{solutionmodel2}),  
for given $\mathcal{K}_1(t)$ and $\mathcal{K}_2(t)$. 
In this way, we illustrate below the crossovers exhibited by the present model 
through the numerical integration of the probability distribution in Eq.~(\ref{solutionmodel2}),
by defining the time-dependent diffusion 
coefficients of Eq. (\ref{segundomodelo}) as
\begin{eqnarray}
\mathcal{K}_1(t) = e^{-\frac{t}{\tau}} \qquad \textnormal{and} \qquad \mathcal{K}_2(t) =1-e^{-\frac{t}{\tau}}~,  
\label{coefficient12}
\end{eqnarray}

\noindent
which satisfy the limits established in Eqs.~(\ref{asymptzero}) and~(\ref{asymptinfty}). 

\begin{figure}[h!]
\centering
\includegraphics[scale=0.70]{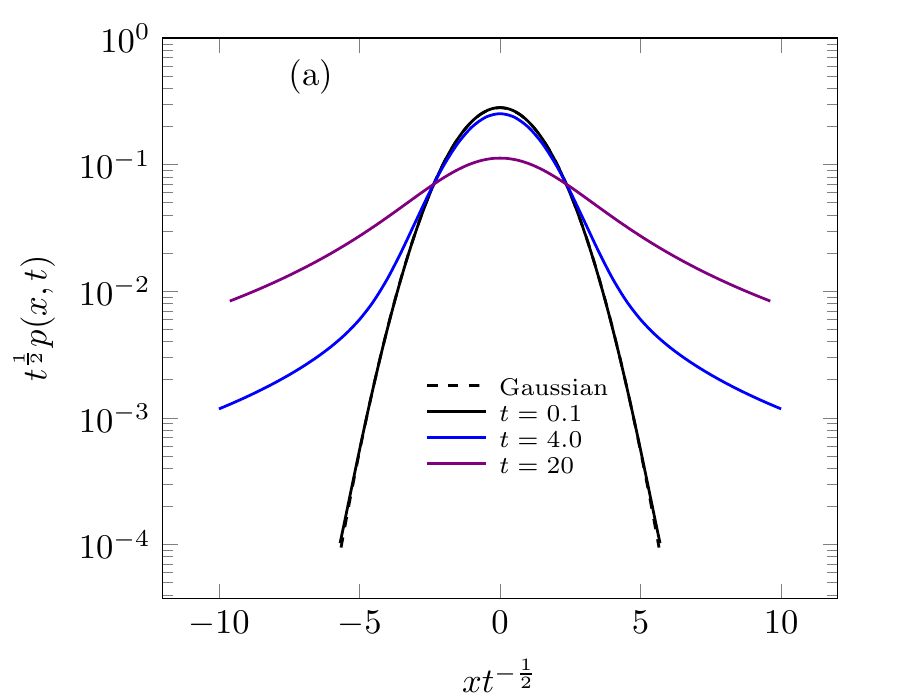} 
\hspace{0.2cm} 
\includegraphics[scale=0.70]{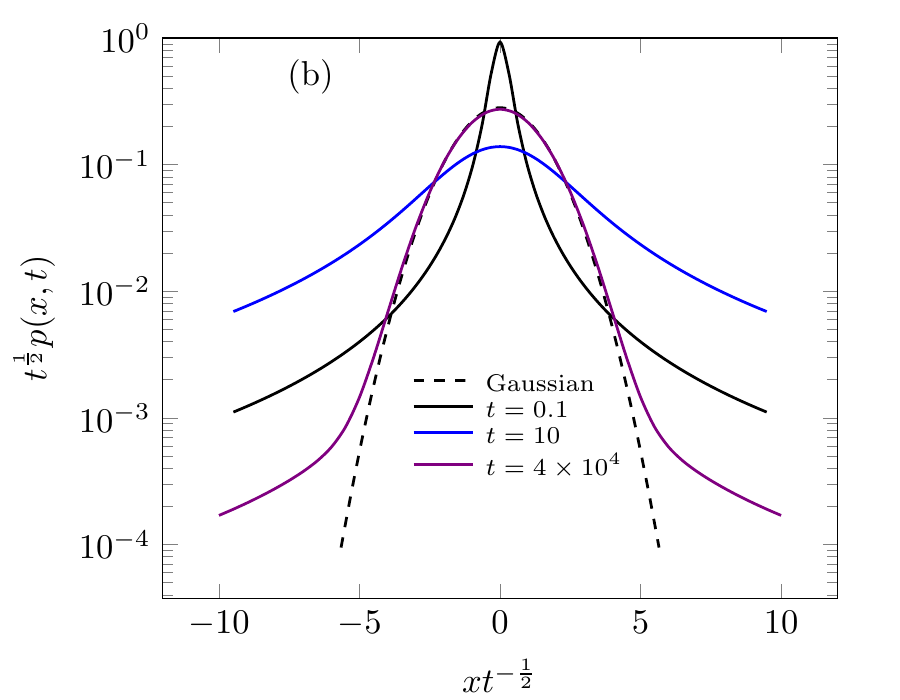} 
\caption{\small{      
The probability distributions $p(x,t)$ are exhibited (using conveniently rescaled variables) 
in log-linear representations, for typical times.
These distributions change in time following the model characterized by 
two different time-dependent diffusion 
coefficients [cf. Eq.~(\ref{segundomodelo})], defined according to 
Eq.~(\ref{coefficient12}), with $\tau=10$.
(a) Time evolution between the Gaussian (represented by a dashed line, superposed
with curve for $t=0.1$) 
and Cauchy distributions, i.e., $\mu_{1}=2$
($t=0$) to $\mu_{2}=1$ $(t \rightarrow \infty)$, showing longer tails for increasing times;
(b) Time evolution between the Cauchy and Gaussian (represented by the 
dashed line) distributions, i.e., $\mu_{1}=1$
($t=0$) to $\mu_{2}=2$ $(t \rightarrow \infty)$, exhibiting 
shorter tails for increasing times.   
}}
\label{fig5}
\end{figure}

 For the same reasons mentioned before in Fig.~\ref{fig2},
we consider the rescaled variables $t^{1/2}p(x,t)$ versus $xt^{-1/2}$ 
in Fig.~\ref{fig5}, which turn out to be convenient in the two Gaussian limits, 
namely, the initial [Fig.~\ref{fig5}(a)],
and long-time distributions [Fig.~\ref{fig5}(b)]. Hence, in 
Fig.~\ref{fig5} we represent $p(x,t)$ (using these conveniently rescaled variables) 
for increasing times, by considering the 
proposal of Eq.~(\ref{segundomodelo}), with the time-dependent diffusion 
coefficients defined according to Eq.~(\ref{coefficient12}).  
These plots were obtained through a numerical 
integration of Eq.~(\ref{solutionmodel2}) for $\tau=10$ and typical values of 
$t$. Two situations are exhibited, namely, crossovers between the Gaussian and 
Cauchy distributions [panel (a)], as well as between the Cauchy and Gaussian [panel (b)]. 
In Fig.~\ref{fig5}(a) 
one notices that deviations from the Gaussian behavior start in the tails, 
whereas the central part persists longer, as shown in the curve for $t=0.4 \tau$;
on the hand, the distribution at time $t=2 \tau$ is essentially inside the crossover 
interval between the two regimes. 
In a similar manner, in Fig.~\ref{fig5}(b) one sees that $p(x,t)$ 
for $t = \tau$ is typically inside the crossover time interval, whereas for 
a much larger time ($t= 4 \times 10^{3} \, \tau$), the 
central part of the distribution has 
essentially converged to the Gaussian, although significant discrepancies persist in 
the tails, requiring even longer times for disappearing.   
This later behavior has already been observed in the previous model, characterized 
by a time-dependent order of the 
fractional derivative, in which the convergence to the final distribution requires longer 
times for accommodating its long tails.
Since both models satisfy the limits established in 
Eqs.~(\ref{asymptzero}) and~(\ref{asymptinfty}), the final 
distribution should be attained after sufficiently long times; however,  
by comparing Figs.~\ref{fig2}(b) and~\ref{fig5}(b) one concludes that the 
disappearance of the tails requires larger times in model 2.  

\begin{figure}[h!]
\centering
\centering
\includegraphics[scale=0.70]{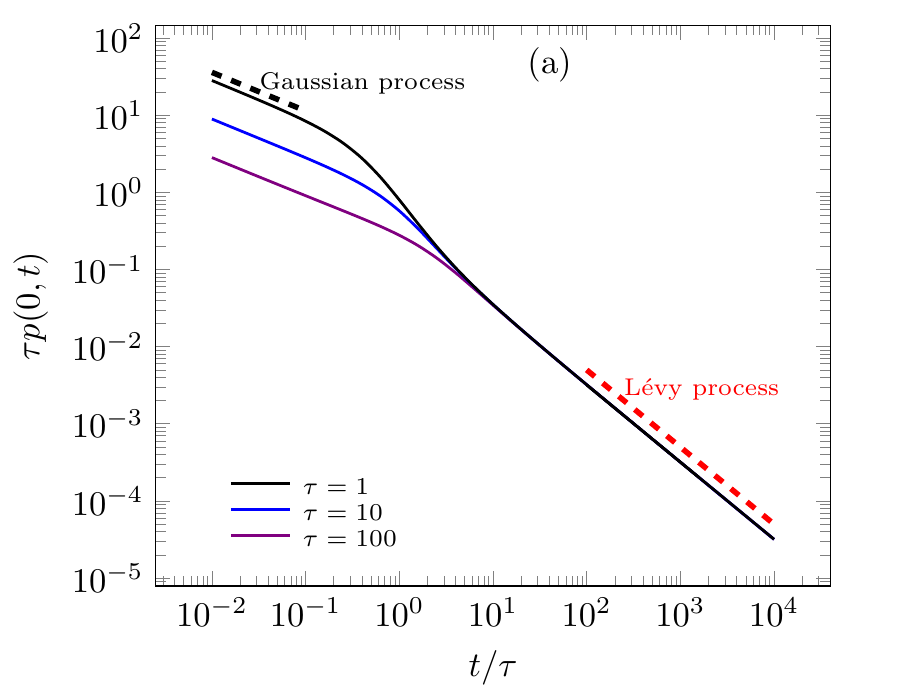} 
\hspace{0.2cm} 
\includegraphics[scale=0.70]{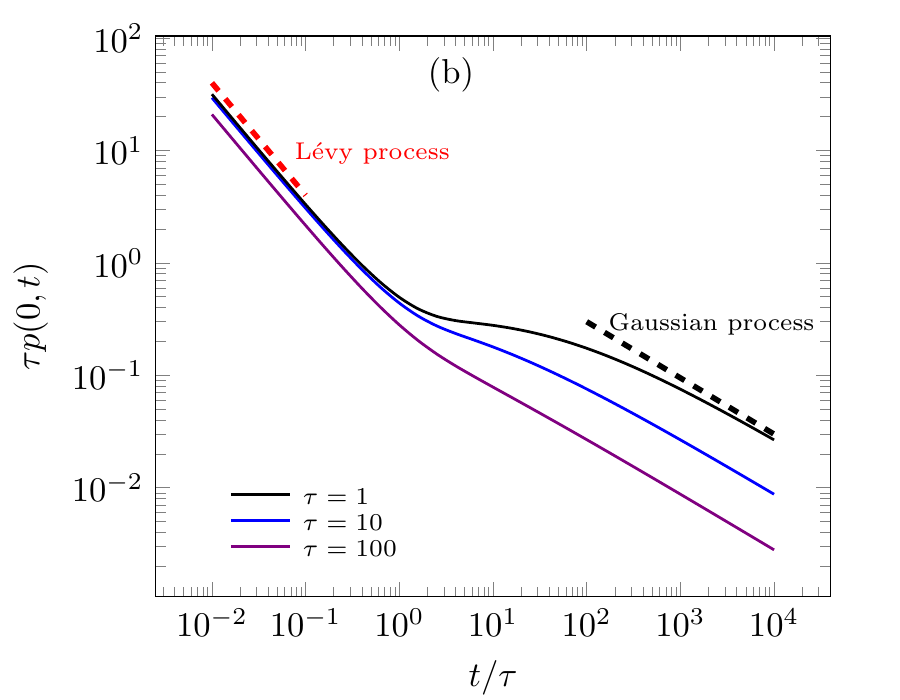} 
\caption{\small{
The return probabilities $p(0,t)$ are plotted versus time (using conveniently rescaled variables) 
in log-log representations, for typical values of $\tau$, following the model 
characterized by two different time-dependent diffusion 
coefficients [cf. Eq.~(\ref{segundomodelo})], defined according to Eq.~(\ref{coefficient12}).   
Away from the crossover region, the scaling of Eq.~(\ref{returnprob}) 
is fulfilled, so that
a crossover between the Gaussian ($\mu=2$) to a Cauchy distribution ($\mu=1$)
is verified in (a), and between a Cauchy and the Gaussian occurs in (b).
In the Gaussian regime, increasing values of $\tau$ apply to curves from top to bottom.
}}
\label{fig6}
\end{figure}

In Fig.~\ref{fig6} we present the return probability $p(0,t)$ versus time in log-log 
representations, considering the model 
defined in Eq.~(\ref{segundomodelo}), with the time-dependent diffusion 
coefficients following Eq.~(\ref{coefficient12}), for typical values of $\tau$; 
these plots were obtained 
by a numerical integration of Eq.~(\ref{solutionmodel2}) 
for $x=0$. The crossover regions occur typically in the time intervals
$10^{-1} < (t/\tau) < 10$ [Fig.~\ref{fig6}(a)] and 
$10^{-1} < (t/\tau) < 10^{2}$ [Fig.~\ref{fig6}(b)], showing that the approach 
to the Gaussian limit requires larger times, in agreement with Fig.~\ref{fig5}(b). 
One notices that away from 
the crossover region, the scaling of Eq.~(\ref{returnprob}) is 
fulfilled, with Fig.~\ref{fig6}(a)
showing a Gaussian regime ($\mu=2$) for $t \ll \tau$, whereas a Cauchy behavior occurs
for $t \gg \tau$; an inverse crossover is observed in Fig.~\ref{fig6}(b).
By comparing Figs.~\ref{fig4}(b) and~\ref{fig6}(b) one sees that 
the present model yields results qualitatively similar to those of model 1, 
i.e., a smooth change between the two regimes, when considering the former model
with $\mu(t)$ following Eq.~(\ref{secondcase}).

One important comment concerns the changes associated with
the distribution tails, like shown in Fig.~\ref{fig5}(b), where
the central part of the distribution has 
converged to the Gaussian, although significant discrepancies persist in 
the tails up to a time $t= 4 \times 10^{3} \, \tau$. 
One should mention that such a behavior, where the central part converges
faster than the tails, is very common in long-tailed-distribution studies (see, 
e.g., Ref~\cite{schwammle2008q}). 
By comparing the largest-time distribution in Fig.~\ref{fig5}(b)
with the corresponding curve in Fig.~\ref{fig6}(b) (cf. case $\tau=10$),  one sees 
that the convergence to the Gaussian limit provided by the analysis of the 
return probability $p(0,t)$ [which occurs for $t \simeq 10^{2} \tau$ in 
Fig.~\ref{fig6}(b)] is incomplete, since this approach focuses only in the time
evolution of the central part of the distribution. In such cases, appropriate
numerical procedures, taking into account the time evolution of the whole distribution
should be carried; as an example where this type of approach has been developed, 
we mention the investigation of the time evolution of $q$-Gaussian distributions, 
as solutions of the porous-medium equation, carried in Ref~\cite{schwammle2008q}.
However, this type of procedure falls out 
of the scope of the present study.  

The model defined in Eq.~(\ref{segundomodelo}), which consists essentially 
in a superposition of two distributions at each time $t$
(except for the limits $t \rightarrow 0$ and $t \rightarrow \infty$), is expected to 
be useful for describing real systems, like  
the diffusion of particles in \textit{cytoskeleton} of a cell~\cite{alencar2016non}. 
Moreover, a superposition of Gaussian and 
L\'evy distributions was investigated in the competition problem 
of two species~\cite{heinsalu2010spatial}. 

\section{Conclusions}
\label{conclusion}

To sum up, we have proposed two models capable of producing crossovers 
among L\'evy flights, characterized by
distinct diffusion exponents. The first one (called model 1) considers a 
time-dependent-order Riesz 
fractional derivative in the  Laplacian term of the diffusion equation, whereas 
the second one (named model 2)
is defined by two fractional-derivative diffusive terms, each of them multiplied by 
different time-dependent diffusion coefficients.   
We have investigated these models by analytical calculations, followed by 
numerical integrations, for obtaining the corresponding distributions $p(x,t)$. 
Although both yield the desired crossovers, attaining the expected limits for 
sufficiently long times, 
the crossovers may occur in qualitatively different ways, due to the fact 
that model 1 is characterized by a single L\'evy process at 
each time $t$, 
with a time-varying exponent $\mu(t)$, whereas in model 2 the distribution 
at time $t$ is expressed in terms of a
convolution of two L\'evy processes, each with its own exponent. 
Typical crossovers among distinct L\'evy processes are illustrated for both cases, 
by following numerically the time evolution of the resulting distribution $p(x,t)$, 
as well as by monitoring the scaling on time of the return 
probability, $p(0,t) \sim t^{-1/\mu}$. 

One should mention that crossovers among distinct types of anomalous diffusion, characterized by different diffusion 
exponents, have been observed often in nature, although proposals of theoretical 
models for describing such effects are very rare in the literature. As an example, one 
has the model introduced recently in Ref.~\cite{sandev2018models},
based on the theory of continuous-time random walks; this model differs from the 
present ones, in the sense that
it leads to a diffusion equation characterized by a time-fractional derivative, whereas in the 
present case (model 1) one has a 
time-dependent-order fractional derivative in the  Laplacian term of the diffusion equation. 
Moreover, a convolution of two L\'evy distributions at a given time $t$, as 
introduced herein in model 2,
is significantly different from the proposal of Ref.~\cite{sandev2018models}.

The crossovers described in the previous section, particularly those between 
L\'evy and Gaussian distributions, have been observed in a class of 
stochastic processes, the so-called truncated L\'evy 
flights~\cite{mantegna1994stochastic}, in econophysical 
applications, like in the behavior of 
commodity-price distributions~\cite{sokolov2004fractional}, as well as 
in the analysis of experimental data in tokamak edge turbulence 
of plasmas~\cite{jhakaw2003}, among many other systems. 

A major question concerns which kind of theoretical model should
be more appropriate for describing a specific crossover observed. 
As expected, both models presented are able to drive systems from 
the initial L\'evy process, 
characterized by an exponent $\mu_{1}$, to the final one, described by an 
exponent $\mu_{2}$. The essential differences between these models appear
in their crossover mechanisms.
In model 1 the crossover occurs through a gradual change in the exponent 
$\mu(t)$, so that in the crossover region the system is expected to approach different
L\'evy regimes, for given time intervals, with $\mu_{1} \leq \mu(t) \leq \mu_{2}$.
In this crossover regime, slow changes in relevant quantities of the environment, 
like temperature, diffusivity, and viscosity, should lead to slightly different
L\'evy flights; therefore, model 1 would be useful to investigate experimental
systems in which different L\'evy processes may occur due to gradual changes 
on environmental properties. On the other hand, a superposition of two Riesz 
operators, considered in model 2, also yields a crossover between two 
distinct L\'evy processes, characterized by exponents $\mu_{1}$ and 
$\mu_{2}$. However, in this case, typical ingredients and properties
of both processes are 
present during the whole crossover interval, so that in different time regimes, 
one L\'evy process becomes more relevant than the other one. 
Differently from the first case, model 2 is expected to present a crossover without 
exploring intermediate L\'evy flights (which may occur between the initial and final 
distributions), but should rather present combinations of the two processes.  

Hence, considering the reasons explained above, one should 
mention that systems characterized by slowly varying properties, like  
light scattering in
a new optical material constructed from glass
microspheres, so-called L\'evy glass~\cite{barthelemy2008levy},
light scattering in atomic vapours~\cite{Mercadier2013}, 
and cold atoms in optical lattices~\cite{Kessler2012} are good candidates for
model 1.  
On the other hand, model 2, which consists essentially in a superposition of 
two distributions at each time $t$, 
is expected to be useful for describing the diffusion of particles in 
\textit{cytoskeleton} of a cell~\cite{alencar2016non}, as 
well as for a two-species competition problem, where a superposition of 
Gaussian and L\'evy distributions has been considered~\cite{heinsalu2010spatial}. 
Additionally,
model 2 should also be relevant in the context of foraging process, 
to investigate random search models that consider a combination of two L\'evy 
processes to study search strategies~\cite{Palyulin2017}.
In general, situations where two species of living beings compete with each other, 
leading to some time of crossover, 
like in food searching, physical-space disputes, among other situations,
become appropriate candidates for model 2.  

In some cases, it may not be obvious to point out a priori which model 
would be more appropriate for 
describing specific applications, and there may be situations where both (or neither) 
of them become suitable. For these, as often occurs with theoretical proposals, only 
through comparison with experimental data, one should be able to choose 
from one of these models. 
However, due to their general characteristics and potential of applicability in 
different contexts, the present proposals should cover many classes of real 
phenomena involving crossovers among distinct L\'evy processes.


\section*{Acknowledgements} 
The authors thank C. Tsallis for fruitful conversations.
M.A.F. dos Santos acknowledges support from the Brazilian agency CAPES (INCT-SC). 
F.D. Nobre and E.M.F. Curado acknowledge partial financial support from CNPq,
CAPES, and FAPERJ (Brazilian funding agencies).  








\end{document}